\documentclass[11 pt]{article}%
\usepackage{amssymb}
\usepackage{graphicx}
\usepackage{amsmath}
\usepackage{amssymb}
\usepackage{amscd}
\usepackage{amsfonts}
\usepackage{cite}%
\providecommand{\U}[1]{\protect\rule{.1in}{.1in}}

 \evensidemargin0in \oddsidemargin0in \topmargin10pt
\RequirePackage{CJK}
\AtBeginDocument{\begin{CJK*}{GBK}{song}\CJKtilde}
\AtEndDocument{\end{CJK*}}

\begin{document}

\title{Entanglement and nonlocality of one- and two-mode combination squeezed
state\thanks{{\small Work supported by a grant from the Key Programs
Foundation of Ministry of Education of China (No. 210115) and the Research
Foundation of the Education Department of Jiangxi Province of China (No.
GJJ10097).}}}
\author{{\small Li-yun Hu}$^{1}${\small \thanks{Corresponding author. E-mail:
hlyun2008@126.com.}, Xue-xiang Xu}$^{1,2}${\small , Qin Guo}$^{1}${\small ,
and Hong-yi Fan}$^{2}$\\$^{1}${\small College of Physics \& Communication Electronics, Jiangxi Normal
University, Nanchang 330022, China}\\$^{2}${\small Department of Physics, Shanghai Jiao Tong University, Shanghai
200030, China}}
\maketitle

\begin{abstract}
{\small We investigate the entanglement and nonlocality properties of one- and
two-mode combination squeezed vacuum state (OTCSS, with two-parameter
}$\lambda$ {\small and }$\gamma${\small ) by analyzing the logarithmic
negativity and the Bell's inequality. It is found that this state exhibits
larger entanglement than that of the usual two-mode squeezed vacuum state
(TSVS), and that in a certain regime of }$\lambda${\small , the violation of
Bell's inequality becomes more obvious, which indicates that the nonlocality
of OTCSS can be stronger than that of TSVS. As an application of OTCSS, the
quantum teleportaion is examined, which shows that there is a region spanned
by }$\lambda$ {\small and }$\gamma${\small in which the fidelity of OTCSS
channel is larger than that of TSVS. }

\end{abstract}


PACS number(s): 42.50.Dv, 03.65.Wj, 03.67.Mn

\section{Introduction}

Entanglement between quantum systems plays a key role in quantum information
processing, such as quantum teleportation, dense coding, and quantum cloning.
In recent years, various entangled states have brought considerable attention
and interests of physicists because of their potential uses in quantum
communication \cite{1,2}. For instance, the two-mode squeezed state is a
typical entangled state of continuous variable and exhibits quantum
entanglement between the idle-mode and the signal-mode in a frequency domain
manifestly. Theoretically, the two-mode squeezed state is constructed by the
two-mode squeezing operator $S=\exp[\lambda(a_{1}a_{2}-a_{1}^{\dagger}%
a_{2}^{\dagger})]$ \cite{3,4,5} acting on the two-mode vacuum state
$\left\vert 00\right\rangle $,%
\begin{equation}
S\left\vert 00\right\rangle =\text{sech}\lambda\exp\left[  -a_{1}^{^{\dagger}%
}a_{2}^{^{\dagger}}\tanh\lambda\right]  \left\vert 00\right\rangle , \label{1}%
\end{equation}
where $\lambda$ is a squeezing parameter, the disentangling of $S$ can be
obtained by using SU(1,1) Lie algebra, $[a_{1}a_{2},a_{1}^{\dagger}%
a_{2}^{\dagger}]=a_{1}^{\dagger}a_{1}+a_{2}^{\dagger}a_{2}+1,$ or by using the
entangled state representation $\left\vert \eta\right\rangle $ \cite{6,7,8},
which was constructed according to the idea of Einstein, Podolsky and Rosen in
their argument that quantum mechanics is incomplete \cite{9}.

Using the relation between Bosonic operators and the coordinate $Q_{i},$
momentum $P_{i},$ $Q_{i}=(a_{i}+a_{i}^{\dagger})/\sqrt{2},\ P_{i}=(a_{i}%
-a_{i}^{\dagger})/(\sqrt{2}\mathtt{i}),$ and introducing the two-mode
quadrature operators of light field, $x_{1}=(Q_{1}+Q_{2})/2,$ $x_{2}%
=(P_{1}+P_{2})/2,$ the variances of $x_{1}$ and $x_{2}$ in the state
$S\left\vert 00\right\rangle $ are in the standard form%
\begin{equation}
\left\langle 00\right\vert S^{\dagger}x_{2}^{2}S\left\vert 00\right\rangle
=\frac{1}{4}e^{-2\lambda},\text{ }\left\langle 00\right\vert S^{\dagger}%
x_{1}^{2}S\left\vert 00\right\rangle =\frac{1}{4}e^{2\lambda}, \label{2}%
\end{equation}
thus we get the standard squeezing for the two quadrature: $x_{1}%
\rightarrow\frac{1}{2}e^{\lambda}x_{1},$ $x_{2}\rightarrow\frac{1}%
{2}e^{-\lambda}x_{2}$. On the other hand, the two-mode squeezing operator can
also be recast into the form $S=\exp\left[  \mathtt{i}\lambda\left(
Q_{1}P_{2}+Q_{2}P_{1}\right)  \right]  .$ Then some interesting questions
naturally rise: what is the property of the following operator%
\begin{equation}
V=\exp\left[  -\mathtt{i}\left(  \lambda_{1}Q_{1}P_{2}+\lambda_{2}Q_{2}%
P_{1}\right)  \right]  , \label{3}%
\end{equation}
with two parameters $\lambda_{1}=\lambda e^{\gamma},\lambda_{2}=\lambda
e^{-\gamma},\lambda>0$? What is the normally ordered expansion of $V$ and what
is the state $V\left\vert 00\right\rangle $? What are the entanglement and
nonlocality properties of $V\left\vert 00\right\rangle ?$ When $\gamma=0,$
Eq.(\ref{3}) just reduces to the usual two-mode squeezing operator $S$. Thus
we can consider $V$ as a generalized two-mode squeezing operator and
$V\left\vert 00\right\rangle $ as one- and two-mode combination squeezed
vacuum state (OTCSS).

In this paper, we investigate entanglement properties and quantum nonlocality
of $V\left\vert 00\right\rangle $ in terms of logarithmic negativity and the
Bell's inequality, respectively. Subsequently, we consider its application in
the field of quantum teleportation by using the characteristic-function
formula. It is shown that this state exhibits larger entanglement than that of
the usual two-mode squeezed vacuum state (TSVS); and in a certain smaller
regime of $\lambda$, that the nonlocality of this state can be stronger than
that of TSVS due to the presence of $\gamma.$ In addition, application to
quantum teleportation with OTCSS is also considered, which shows that there is
a region spanned by $\lambda$ and $\gamma$ in which the fidelity of OTCSS
channel is larger than that of TSVS.

Our paper is arranged as follows. In section 2, we derive the normal ordering
form of one- and two-mode combination squeezing operator by using the
technique of integration within an ordered product (IWOP) of operators. In
section 3, using the Weyl ordering form of single-mode Wigner operator and the
order-invariance of Weyl ordered operators under similar transformations, we
derive analytically the Wigner function of $V\left\vert 00\right\rangle $.
Sections 4 and 5 are devoted to investigating the entanglement properties and
the nonlocal properties OTCSS by using the Bell's inequality and the
logarithmic negativity, respectively.\ An application to quantum teleportation
with OTCSS is involved in section 6. We end with the main conclusions of our work.

\section{The normal ordering form of $V$ and fluctuations in $V\left\vert
00\right\rangle $}

In order to know $V\left\vert 00\right\rangle ,$ we need to derive the normal
ordering form of the unitary operator $V$ by virtue of the IWOP technique
\cite{10,11,12}. Using the Baker-Hausdorff formula,
\begin{equation}
e^{A}Be^{-A}=B+\left[  A,B\right]  +\frac{1}{2!}\left[  A,\left[  A,B\right]
\right]  +\frac{1}{3!}\left[  A,\left[  A,\left[  A,B\right]  \right]
\right]  +\cdots, \label{4}%
\end{equation}
and noticing that%
\begin{align}
i\left[  \lambda_{1}Q_{1}P_{2}+\lambda_{2}Q_{2}P_{1},Q_{1}\right]   &
=\lambda_{2}Q_{2},\text{ }\label{5}\\
i\left[  \lambda_{1}Q_{1}P_{2}+\lambda_{2}Q_{2}P_{1},Q_{2}\right]   &
=\lambda_{1}Q_{1},\\
i\left[  \lambda_{1}Q_{1}P_{2}+\lambda_{2}Q_{2}P_{1},P_{1}\right]   &
=-\lambda_{1}P_{2},\\
i\left[  \lambda_{1}Q_{1}P_{2}+\lambda_{2}Q_{2}P_{1},P_{2}\right]   &
=-\lambda_{2}P_{1}, \label{6}%
\end{align}
we have%
\begin{align}
V^{-1}Q_{1}V  &  =Q_{1}\cosh\lambda+Q_{2}e^{-\gamma}\sinh\lambda,\label{7}\\
V^{-1}Q_{2}V  &  =Q_{2}\cosh\lambda+Q_{1}e^{\gamma}\sinh\lambda,\\
V^{-1}P_{1}V  &  =P_{1}\cosh\lambda-P_{2}e^{\gamma}\sinh\lambda,\\
V^{-1}P_{2}V  &  =P_{2}\cosh\lambda-P_{1}e^{-\gamma}\sinh\lambda. \label{8}%
\end{align}
Thus, in order to keep the eigenvalues invariant under the $V$ transformation,
i..e.,
\begin{equation}
V^{-1}Q_{k}V\left\vert q_{1}q_{2}\right\rangle ^{\prime}=q_{k}\left\vert
q_{1}q_{2}\right\rangle ^{\prime},(k=1,2), \label{9}%
\end{equation}
the base vector must be changed to%
\begin{equation}
\left\vert q_{1}q_{2}\right\rangle ^{\prime}=V^{-1}\left\vert q_{1}%
q_{2}\right\rangle =\left\vert \Lambda^{-1}\left(
\begin{array}
[c]{c}%
q_{1}\\
q_{2}%
\end{array}
\right)  \right\rangle ,\Lambda=\left(
\begin{array}
[c]{cc}%
\cosh\lambda & e^{-\gamma}\sinh\lambda\\
e^{\gamma}\sinh\lambda & \cosh\lambda
\end{array}
\right)  , \label{10}%
\end{equation}
where $\left\vert q_{1}q_{2}\right\rangle =\left\vert q_{1}\right\rangle
\otimes\left\vert q_{2}\right\rangle $, and $\left\vert q_{k}\right\rangle $
is the coordinate eigenstate,
\begin{equation}
\left\vert q_{k}\right\rangle =\pi^{-1/4}\exp[-\frac{1}{2}q^{2}+\sqrt
{2}qa^{\dag}-\frac{1}{2}a^{\dag2}]\left\vert 0\right\rangle . \label{11}%
\end{equation}

Using the completeness raltion $\int_{-\infty}^{\infty}dq_{1}dq_{2}\left\vert
q_{1},q_{2}\right\rangle \left\langle q_{1},q_{2}\right\vert =1$, we have%
\begin{equation}
V^{-1}=\int_{-\infty}^{\infty}dq_{1}dq_{2}\left\vert \Lambda^{-1}\left(
\begin{array}
[c]{c}%
q_{1}\\
q_{2}%
\end{array}
\right)  \right\rangle \left\langle q_{1},q_{2}\right\vert , \label{12}%
\end{equation}
which leads to
\begin{equation}
V=\int_{-\infty}^{\infty}dq_{1}dq_{2}\left\vert \Lambda\left(
\begin{array}
[c]{c}%
q_{1}\\
q_{2}%
\end{array}
\right)  \right\rangle \left\langle q_{1},q_{2}\right\vert . \label{13}%
\end{equation}
Actually, one can check (\ref{13}) by $V^{-1}V=VV^{-1}=1.$ Further using the
vacuum projector $\left\vert 00\right\rangle \left\langle 00\right\vert
=\colon\exp[-a^{\dag}a-b^{\dag}b]\colon(\colon\colon$ denoting normal
ordering$)$, as well as the IWOP technique, we can put $V$ into the normal
ordering form \cite{13},%
\begin{align}
V  &  =\frac{2}{\sqrt{L}}\exp\left\{  \frac{1}{L}\left[  \left(  b^{\dag
2}-a^{\dag2}\right)  \sinh^{2}\lambda\sinh2\gamma+2a^{\dag}b^{\dag}%
\sinh2\lambda\cosh\gamma\right]  \right\} \nonumber\\
&  \colon\exp\left\{  \frac{4}{L}\left[  \left(  a^{\dag}a+b^{\dag}b\right)
\cosh\lambda+\left(  b^{\dag}a-a^{\dag}b\right)  \sinh\lambda\sinh
\gamma\right]  -a^{\dag}a-b^{\dag}b\right\}  \colon\nonumber\\
&  \exp\left\{  \frac{1}{L}\left[  \left(  b^{2}-a^{2}\right)  \sinh
^{2}\lambda\sinh2\gamma-2a^{\dag}b^{\dag}\sinh2\lambda\cosh\gamma\right]
\right\}  , \label{14}%
\end{align}
where $L=4\left(  1+\sinh^{2}\gamma\tanh^{2}\lambda\right)  \cosh^{2}\lambda.$
Eq. (\ref{14}) is just the normal ordering form of $V$. It is obviously to see
that when $\gamma=0$, Eq.(\ref{14}) just reduces to the usual two-mode
squeezing operator. Operating $V$ on the two-mode vacuum state $\left\vert
00\right\rangle $, we obtain the squeezed vacuum state,
\begin{equation}
V\left\vert 00\right\rangle =\frac{2}{\sqrt{L}}\exp\left\{  \frac{1}{L}\left[
\left(  b^{\dag2}-a^{\dag2}\right)  \sinh^{2}\lambda\sinh2\gamma+2a^{\dag
}b^{\dag}\sinh2\lambda\cosh\gamma\right]  \right\}  \left\vert 00\right\rangle
. \label{15}%
\end{equation}

On the other hand, by using the transformations Eqs.(\ref{7})-(\ref{8}), one
can derive the variances of $x_{1}$ and $x_{2}$ in the state $V\left\vert
00\right\rangle $ \cite{13}
\begin{align}
\left\langle \left(  \Delta x_{1}\right)  ^{2}\right\rangle  &  =\frac{1}%
{4}\left(  \cosh2\lambda+2\sinh^{2}\lambda\sinh^{2}\gamma+\sinh2\lambda
\cosh\gamma\right)  ,\label{16}\\
\left\langle \left(  \Delta x_{2}\right)  ^{2}\right\rangle  &  =\frac{1}%
{4}\left(  \cosh2\lambda+2\sinh^{2}\lambda\sinh^{2}\gamma-\sinh2\lambda
\cosh\gamma\right)  , \label{17}%
\end{align}
which indicate that the variances are not only dependent on parameter
$\lambda$, but also on parameter $\gamma.$ When $\gamma=0,$ Eqs.(\ref{16}) and
(\ref{17}) reduce to $\left\langle \left(  \Delta x_{1}\right)  ^{2}%
\right\rangle =\frac{1}{4}e^{2\lambda},$ and $\left\langle \left(  \Delta
x_{2}\right)  ^{2}\right\rangle =\frac{1}{4}e^{-2\lambda},$ corresponding to
the usual TSVS. In particular, by modulating the two parameters ($\lambda$ and
$\gamma$), we can realize that%
\begin{equation}
\left\langle \left(  \Delta x_{1}\right)  ^{2}\right\rangle >\frac{1}%
{4}e^{2\lambda},\left\langle \left(  \Delta x_{2}\right)  ^{2}\right\rangle
<\frac{1}{4}e^{-2\lambda}, \label{18}%
\end{equation}
whose condition is given by
\begin{equation}
0<\tanh\lambda<\frac{1}{1+\cosh\gamma},\lambda>0, \label{19}%
\end{equation}
which mean that the OTCSS may exhibit stronger squeezing in one quadrature
than that of the TSVS while exhibiting weaker squeezing in another quadrature
when the condition (\ref{19}) satisfied. Then,\ can the OTCSS exhibits
stronger nonlocality or more observable violation of Bell's inequality? In the
following, we pay our attention to these two aspects.

\section{Wigner function of $V\left\vert 00\right\rangle $}

Wigner distribution functions \cite{14,15,16} of quantum states are widely
studied in quantum statistics and quantum optics. Now we derive the expression
of the Wigner function of $V\left\vert 00\right\rangle .$ Here we take a new
method to do it. Recalling that in Ref.\cite{17,18,19} we have introduced the
Weyl ordering form of single-mode Wigner operator $\Delta_{1}\left(
q_{1},p_{1}\right)  $,%
\begin{equation}
\Delta_{1}\left(  q_{1},p_{1}\right)  =%
\genfrac{}{}{0pt}{}{:}{:}%
\delta\left(  q_{1}-Q_{1}\right)  \delta\left(  p_{1}-P_{1}\right)
\genfrac{}{}{0pt}{}{:}{:}%
, \label{20}%
\end{equation}
its normal ordering form is%
\begin{equation}
\Delta_{1}\left(  q_{1},p_{1}\right)  =\frac{1}{\pi}\colon\exp\left[  -\left(
q_{1}-Q_{1}\right)  ^{2}-\left(  p_{1}-P_{1}\right)  ^{2}\right]  \colon,
\label{21}%
\end{equation}
where the symbols $\colon\colon$ and $%
\genfrac{}{}{0pt}{}{:}{:}%
\genfrac{}{}{0pt}{}{:}{:}%
$ denote the normal ordering and the Weyl ordering, respectively. Note that
the order of Bose operators $a_{1}$ and $a_{1}^{\dagger}$ within a normally
ordered product and a Weyl ordered product can be permuted. That is to say,
even though $[a_{1},a_{1}^{\dagger}]=1$, we can have $\colon a_{1}%
a_{1}^{\dagger}\colon=\colon a_{1}^{\dagger}a_{1}\colon$ and$%
\genfrac{}{}{0pt}{}{:}{:}%
a_{1}a_{1}^{\dagger}%
\genfrac{}{}{0pt}{}{:}{:}%
=%
\genfrac{}{}{0pt}{}{:}{:}%
a_{1}^{\dagger}a_{1}%
\genfrac{}{}{0pt}{}{:}{:}%
.$

For one- and two-mode combination squeezed vacuum state $V\left\vert
00\right\rangle $, its Wigner function is given by%
\begin{equation}
W\left(  q_{1},p_{1};q_{2},p_{2}\right)  =\mathtt{tr}\left[  V\left\vert
00\right\rangle \left\langle 00\right\vert V^{-1}\Delta_{1}\left(  q_{1}%
,p_{1}\right)  \Delta_{2}\left(  q_{2},p_{2}\right)  \right]  =\left\langle
00\right\vert U\left\vert 00\right\rangle , \label{22}%
\end{equation}
where $U=V^{-1}\Delta_{1}\left(  q_{1},p_{1}\right)  \Delta_{2}\left(
q_{2},p_{2}\right)  V.$ Further using Eq.(\ref{20}), noticing that the Weyl
ordering has a remarkable property, i.e., the order-invariance of Weyl ordered
operators under similar transformations \cite{17,18,19}, which means%
\begin{equation}
V^{-1}%
\genfrac{}{}{0pt}{}{:}{:}%
\left(  \circ\circ\circ\right)
\genfrac{}{}{0pt}{}{:}{:}%
V=%
\genfrac{}{}{0pt}{}{:}{:}%
V^{-1}\left(  \circ\circ\circ\right)  V%
\genfrac{}{}{0pt}{}{:}{:}%
, \label{23}%
\end{equation}
as if the \textquotedblleft fence" $%
\genfrac{}{}{0pt}{}{:}{:}%
\genfrac{}{}{0pt}{}{:}{:}%
$did not exist, thus $U$\ can be cast into the following form (see appendix
A),%
\begin{align}
U  &  =V^{-1}%
\genfrac{}{}{0pt}{}{:}{:}%
\delta\left(  q_{1}-Q_{1}\right)  \delta\left(  p_{1}-P_{1}\right)
\delta\left(  q_{2}-Q_{2}\right)  \delta\left(  p_{2}-P_{2}\right)
\genfrac{}{}{0pt}{}{:}{:}%
V\nonumber\\
&  =\Delta_{1}\left(  q_{1}\cosh\lambda-q_{2}e^{-\gamma}\sinh\lambda
,p_{1}\cosh\lambda+p_{2}e^{\gamma}\sinh\lambda\right) \nonumber\\
&  \times\Delta_{2}\left(  q_{2}\cosh\lambda-q_{1}e^{\gamma}\sinh\lambda
,p_{2}\cosh\lambda+p_{1}e^{-\gamma}\sinh\lambda\right)  . \label{24}%
\end{align}
So the Wigner function of $V\left\vert 00\right\rangle $ is given by
\begin{equation}
W\left(  q_{1},p_{1};q_{2},p_{2}\right)  =\frac{1}{\pi^{2}}\exp\left\{
-m_{1}\left(  q_{1}^{2}+p_{2}^{2}\right)  \allowbreak-m_{2}\left(  p_{1}%
^{2}+q_{2}^{2}\right)  +2\left(  q_{1}q_{2}\allowbreak-p_{1}p_{2}\right)
m_{3}\right\}  , \label{25}%
\end{equation}
where
\[
m_{1}=\cosh^{2}\lambda+e^{2\gamma}\sinh^{2}\lambda,\text{ }m_{2}=\cosh
^{2}\lambda+e^{-2\gamma}\allowbreak\sinh^{2}\lambda,\text{ }m_{3}=\cosh
\gamma\sinh2\lambda.
\]
In particular, when $\gamma=0$, Eq.(\ref{25}) becomes%
\begin{align}
W\left(  q_{1},p_{1};q_{2},p_{2}\right)   &  =\frac{1}{\pi^{2}}\exp\left\{
-\left(  p_{1}^{2}+p_{2}^{2}+q_{1}^{2}+q_{2}^{2}\right)  \cosh2\lambda
\allowbreak\right. \nonumber\\
&  \left.  +2\left(  q_{1}q_{2}-p_{1}p_{2}\right)  \sinh2\lambda\right\}  ,
\label{26}%
\end{align}
which is just the Wigner function of the usual TSVS.

\section{Entanglement properties of $V\left\vert 00\right\rangle $}

In this section, we consider the entanglement properties of $V\left\vert
00\right\rangle $. It is well known that a two-mode Gaussian state can be
completely characterized by its first and second statistical moments and the
covariance matrix of elements $\sigma$. In general, the first statistical
moments can be adjusted by local displacements without affecting entanglement,
thus they they will can be set to be zero without loss of generality and the
behavior of the covariance matrix $\sigma$ is all important for the study of
entanglement. There are several quantitative measurements of quantum
entanglement proposed \cite{20,21,22}. For a two-mode Gaussian state, the
entanglement is best characterized by the logarithmic negativity
$E_{\mathcal{N}}$, a quantity evaluated in terms of the symplectic eigenvalues
of $\sigma$ \cite{23,24}.

In order to evaluate the entanglement of $V\left\vert 00\right\rangle ,$ we
reform Eq.(\ref{25}) as follows in terms of phase space quadrature variables,%

\begin{equation}
W\left(  q_{1},p_{1};q_{2},p_{2}\right)  =\frac{1}{\pi^{2}}\exp\left[
-\frac{1}{2}\left(
\begin{array}
[c]{cccc}%
q_{1} & p_{1} & q_{2} & p_{2}%
\end{array}
\right)  \sigma^{-1}\left(
\begin{array}
[c]{cccc}%
q_{1} & p_{1} & q_{2} & p_{2}%
\end{array}
\right)  ^{T}\right]  , \label{27}%
\end{equation}
where the covariance matrix $\sigma$ of this OTCSS is \cite{25}
\begin{equation}
\sigma=\left(
\begin{array}
[c]{cc}%
u & w\\
w^{T} & v
\end{array}
\right)  ,\text{ }u=\frac{1}{2}\left(
\begin{array}
[c]{cc}%
m_{2} & 0\\
0 & m_{1}%
\end{array}
\right)  ,v=\frac{1}{2}\left(
\begin{array}
[c]{cc}%
m_{1} & 0\\
0 & m_{2}%
\end{array}
\right)  ,w=\frac{1}{2}\left(
\begin{array}
[c]{cc}%
m_{3} & 0\\
0 & -m_{3}%
\end{array}
\right)  . \label{28}%
\end{equation}
In particular, when $\gamma=0$, $m_{1}=m_{2}=\cosh2\lambda,$ $m_{3}%
=\sinh2\lambda$, Eq.(\ref{28}) just reduces to the so-called standard form of
covariance matrix for TSVS \cite{24}.

The condition for entanglement of a Gaussian state is derived from the
partially transposed density matrix (PPT criterion) \cite{24}, according to
the smallest symplectic eigenvalue $\tilde{n}_{s}$ of the partially transposed
state, $\tilde{n}_{s}<\frac{1}{2},$ i.e., $\tilde{n}_{s}\geqslant\frac{1}{2}$
means the a two-mode Gaussian state is separable, where $\tilde{n}_{s}$ is
defined as
\begin{equation}
\tilde{n}_{s}=\min\left[  \tilde{n}_{+},\tilde{n}_{-}\right]  , \label{30}%
\end{equation}
and $\tilde{n}_{\pm}\ $is given by \cite{26}%
\begin{equation}
\tilde{n}_{\pm}=\sqrt{\frac{\tilde{\Delta}\left(  \sigma\right)  \pm
(\tilde{\Delta}\left(  \sigma\right)  ^{2}-4\det\sigma)^{1/2}}{2}}, \label{31}%
\end{equation}
where $\tilde{\Delta}\left(  \sigma\right)  =\Delta\left(  \tilde{\sigma
}\right)  =\det u+\det v-2\det w.$

Using Eqs.(\ref{28})-(\ref{31}), the corresponding sympletic eigenvalues
$\tilde{n}_{\pm}$ are then given by
\begin{equation}
\tilde{n}_{\pm}=\frac{1}{2}\left(  \sqrt{m_{1}m_{2}}\pm m_{3}\right)  .
\label{32}%
\end{equation}
One the other hand, the corresponding quantification of entanglement is given
by the logarithmic negativity $E_{\mathcal{N}}$ defined as \cite{20,26,27},%
\begin{equation}
E_{\mathcal{N}}=\max\left[  0,-\ln2\tilde{n}_{s}\right]  . \label{33}%
\end{equation}
From Eqs.(\ref{28}), (\ref{32}) and (\ref{33}), one can clearly see that the
logarithmic negativity $E_{\mathcal{N}}$ is dependent on $\lambda$ and
$\gamma.$ In figure 1, we plot the logarithmic negativity $E_{\mathcal{N}}$ as
a function of parameters $\lambda$ and $\gamma$. From Fig.1, we clearly see a
new feature, i.e., in presence of parameter $\gamma$, the logarithmic
negativity becomes larger than that of tha usual squeezed state ($\gamma=0$).

\begin{figure}[ptb]
\label{Fig1}
\centering\includegraphics[width=8cm]{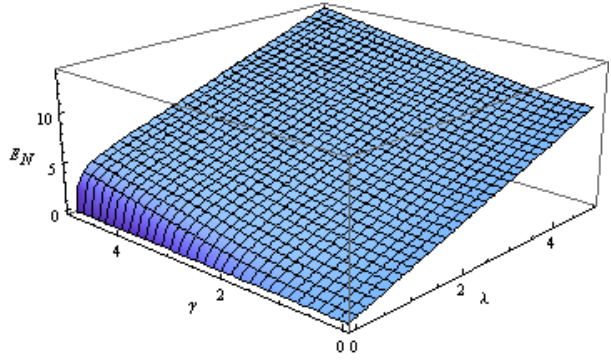}\caption{{\protect\small (Color
online) The logarithmic negativity }$E_{\mathcal{N}}${\protect\small as a
function of parameters }$\lambda${\protect\small and }$\gamma$%
{\protect\small .}}%
\end{figure}

\section{Violations of Bell's inequality for $V\left\vert 00\right\rangle $}

We now turn our attention to the nonlocal properties of $V\left\vert
00\right\rangle $\ in terms of the Bell's inequality. For a two-mode
continuous variable system, the Bell's inequality is given, using correlations
between parity measurement, by
\begin{equation}
\left\vert \mathcal{B}\right\vert \equiv\left\vert \left\langle \hat{\Pi}%
^{ab}\left(  \alpha^{\prime},\beta^{\prime}\right)  +\hat{\Pi}^{ab}\left(
\alpha,\beta^{\prime}\right)  +\hat{\Pi}^{ab}\left(  \alpha^{\prime}%
,\beta\right)  -\hat{\Pi}^{ab}\left(  \alpha,\beta\right)  \right\rangle
\right\vert \leqslant2, \label{34}%
\end{equation}
where $B$ is the Bell function, and the superscripts $a$ and $b$ denote the
modes and $\hat{\Pi}^{ab}\left(  \alpha,\beta\right)  $ is the displaced
parity operator (Wigner operator)\cite{28} defined as
\begin{align}
\hat{\Pi}^{ab}\left(  \alpha,\beta\right)   &  \equiv\hat{\Pi}^{a}\left(
\alpha\right)  \hat{\Pi}^{b}\left(  \beta\right)  =D\left(  \alpha\right)
D\left(  \beta\right)  (-1)^{a^{\dag}a+b^{\dag}b}D^{\dag}\left(  \beta\right)
D^{\dag}\left(  \alpha\right) \nonumber\\
&  =\pi^{2}\Delta_{a}\left(  \alpha\right)  \Delta_{b}\left(  \beta\right)  ,
\label{35}%
\end{align}
where $\alpha=(q_{1}+ip_{1})/\sqrt{2},\beta=(q_{2}+ip_{2})/\sqrt{2}.$ The
expectation value of this displaced parity operator is just proportional to
the two-mode Wigner function, i.e.,
\begin{equation}
\Pi\left(  \alpha,\beta\right)  =\mathtt{tr}\left[  \rho\hat{\Pi}^{ab}\left(
\alpha,\beta\right)  \right]  =\pi^{2}W\left(  \alpha,\beta\right)  ,
\label{36}%
\end{equation}
which shows that the connection between this displaced parity operator and
Wigner function provides an equivalent definition \cite{29}.

The Bell function is measured for any of four combinations of $\alpha
=0,\sqrt{J}e^{i\varphi}$ and $\beta=0,\sqrt{J}e^{i\theta}$, where
$J(=\left\vert \alpha\right\vert ^{2}=\left\vert \beta\right\vert ^{2})$ is a
positive constant characterizing the magnitude of the displacement. From these
quantities we construct the combination \cite{30}
\begin{equation}
\mathcal{B}\equiv\pi^{2}\left[  W\left(  0,0\right)  +W\left(  \sqrt
{J}e^{i\varphi},0\right)  +W\left(  0,\sqrt{J}e^{i\theta}\right)  -W\left(
\sqrt{J}e^{i\varphi},\sqrt{J}e^{i\theta}\right)  \right]  . \label{37}%
\end{equation}
In particular, when $\varphi=0$, $\theta=\pi,$ Eq.(\ref{37}) just reduces to
Eq.(7) in Ref.\cite{28}. Then we can test Bell's inequality $-2\leqslant
\mathcal{B}\leqslant2$ by means of the two-mode Wigner function measurement.
Recently, a generalized quasiprobability function is proposed to test quantum
nonlocality \cite{31}, which includes two-type of Bell-inequality by using the
Wigner function \cite{32} and the $Q-$function \cite{30} as its limiting cases.

By noticing that $\alpha=(q_{1}+ip_{1})/\sqrt{2},\beta=(q_{2}+ip_{2})/\sqrt
{2}$, and $\left(
\begin{array}
[c]{cccc}%
q_{1} & p_{1} & q_{2} & p_{2}%
\end{array}
\right)  N^{-1}=\left(
\begin{array}
[c]{cccc}%
\alpha^{\ast} & \alpha & \beta^{\ast} & \beta
\end{array}
\right)  ,N=\frac{1}{\sqrt{2}}\left(
\begin{array}
[c]{cccc}%
1 & i & 0 & 0\\
1 & -i & 0 & 0\\
0 & 0 & 1 & i\\
0 & 0 & 1 & -i
\end{array}
\right)  ,$ we can put Eq.(\ref{27}) into another form
\begin{equation}
W\left(  \alpha;\beta\right)  =\frac{1}{\pi^{2}}\exp\left[  -\frac{1}%
{2}\left(
\begin{array}
[c]{cccc}%
\alpha^{\ast} & \alpha & \beta^{\ast} & \beta
\end{array}
\right)  M\left(
\begin{array}
[c]{cccc}%
\alpha^{\ast} & \alpha & \beta^{\ast} & \beta
\end{array}
\right)  ^{T}\right]  , \label{37b}%
\end{equation}
where $\bar{M}$ is a $4\times4$ Hermitian matrix
\begin{equation}
M=\left(
\begin{array}
[c]{cccc}%
m_{1}-m_{2} & m_{1}+m_{2} & -2m_{3} & 0\\
m_{1}+m_{2} & m_{1}-m_{2} & 0 & -2m_{3}\\
-2m_{3} & 0 & m_{2}-m_{1} & m_{1}+m_{2}\\
0 & -2m_{3} & m_{1}+m_{2} & m_{2}-m_{1}%
\end{array}
\right)  . \label{37c}%
\end{equation}

Substituting Eq.(\ref{37b}) into Eq.(\ref{37}) we have%
\begin{align}
\mathcal{B}  &  =1+\exp\left[  -2J\cosh^{2}\lambda-2J\left(  e^{2\gamma}%
\cos^{2}\varphi+e^{-2\gamma}\sin^{2}\varphi\allowbreak\right)  \sinh
^{2}\lambda\right] \nonumber\\
&  \text{ \ \ \ \ }+\exp\left[  -2J\cosh^{2}\lambda-2J\left(  e^{2\gamma}%
\sin^{2}\theta+e^{-2\gamma}\cos^{2}\theta\allowbreak\right)  \sinh^{2}%
\lambda\right] \nonumber\\
&  \text{ \ \ \ \ }-\exp\left\{  -4J\cosh^{2}\lambda-2J\left(  \cos^{2}%
\varphi+\sin^{2}\theta\right)  e^{2\gamma}\sinh^{2}\lambda\right. \nonumber\\
\text{ \ \ \ \ \ }  &  \ \ \ \ \ \ \ \left.  -2J\left(  \sin^{2}\varphi
+\cos^{2}\theta\right)  e^{-2\gamma}\allowbreak\sinh^{2}\lambda+4J\cos\left(
\theta+\varphi\right)  \cosh\gamma\sinh2\lambda\right\}  , \label{38}%
\end{align}
Thus we can say that $V\left\vert 00\right\rangle $ is quantum mechanically
nonlocal as $\left\vert \mathcal{B}\right\vert >2,$ and the nonlocality is
stronger with the increase of $\left\vert \mathcal{B}\right\vert $. From
Eq.(\ref{38}) one can see that the degree of nonlocality not only depends on
the coherent amplitude $J$, on the phases $\theta$ and $\varphi$, but also on
the parameter $\gamma$. In particular, when $\varphi=0$, $\theta=\pi,$
Eq.(\ref{38}) just reduces to
\begin{align}
\mathcal{B}  &  =1+\exp\left[  -2J\left(  \cosh^{2}\lambda+e^{2\gamma}%
\sinh^{2}\lambda\right)  \right] \nonumber\\
&  \text{ \ \ \ \ }+\exp\left[  -2J\left(  \cosh^{2}\lambda+e^{-2\gamma}%
\sinh^{2}\lambda\right)  \right] \nonumber\\
&  \text{ \ \ \ \ }-\exp\left[  -4J\left(  \cosh^{2}\lambda+\cosh2\gamma
\sinh^{2}\lambda\right)  -4J\allowbreak\cosh\gamma\sinh2\lambda\right]  ,
\label{39}%
\end{align}
which further becomes Eq.(7) in Ref.\cite{28} with $\gamma=0$.

\begin{figure}[ptb]
\label{Fig2}
\centering\includegraphics[width=8cm]{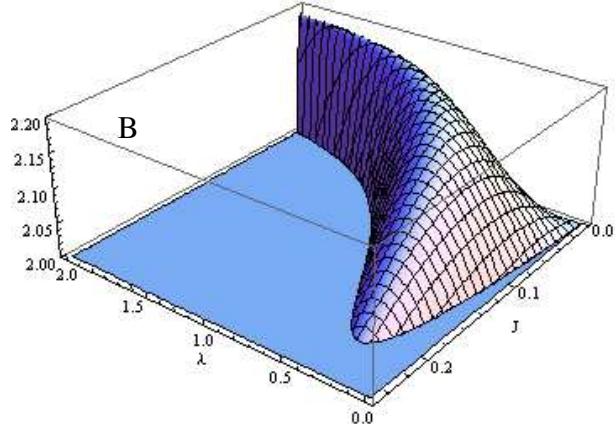}\caption{{\protect\small (Color
online) Plot of the Bell function }$\mathcal{B}${\protect\small as a function
of parameters }$\lambda${\protect\small and }$J$, {\protect\small for }%
$\gamma=0,\theta=\pi,\varphi=0${\protect\small . Only values exceeding the
bound imposed by local theories are shown.}}%
\end{figure}

In figure 2, we plot Bell function in the space spanned by parameters $J$ and
$\lambda$ with $\gamma=0$ (corresponding to the usual squeezed vacuum state).
From Fig. 2, one can clearly see that the result (\ref{38}) violates the upper
bound imposed by local theories. With the increase of $\lambda$, the violation
of Bell's inequality becomes more observable for smaller $J$.
\begin{figure}[ptb]
\label{Fig3}
\centering\includegraphics[width=8cm]{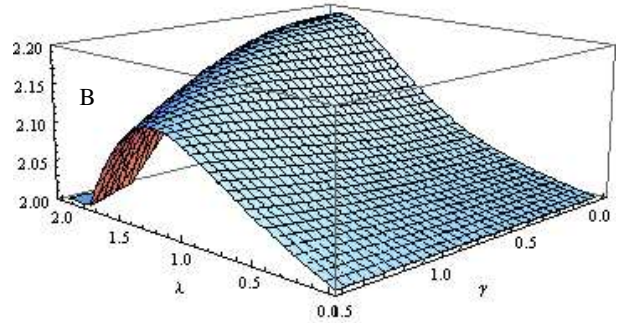}\caption{{\protect\small (Color
online) Plot of the Bell function }$\mathcal{B}${\protect\small as a function
of parameters }$\lambda$ {\protect\small and }$\gamma,${\protect\small for
given }$\left\vert \alpha\right\vert =\left\vert \beta\right\vert
=0.05${\protect\small and }$\theta=\pi,\varphi=0.${\protect\small Only values
exceeding the bound imposed by local theories are shown.}}%
\end{figure}\begin{figure}[ptb]
\label{Fig4}
\centering\includegraphics[width=8cm]{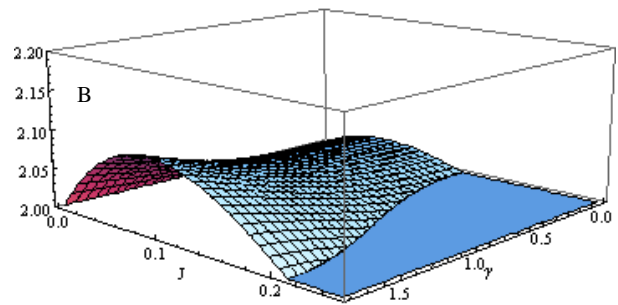}\caption{{\protect\small (Color
online) Plot of the Bell function }$\mathcal{B}${\protect\small as a function
of parameters }$J$ {\protect\small and }$\gamma$,{\protect\small for given
}$\theta=\pi,\varphi=0,${\protect\small and }$\lambda=0.1${\protect\small .
Only values exceeding the bound imposed by local theories are shown.}}%
\end{figure}

As depicted in Fig. 3, the Bell function also violates the upper bound in the
space spanned by parameters $\gamma$ and $\lambda$ with given $\alpha,\beta$
values. From Fig.3, one can see that for a given small $\gamma,$ the violation
of Bell's inequality becomes more observable with the increase of $\lambda$;
while for a large $\gamma$, the Bell function is not always monotone for an
increasing $\lambda$; In a certain smaller regime of $\lambda$, it is found
that the violation of Bell's inequality becomes more observable with
increasing $\gamma$, which indicates that the nonlocality of $V\left\vert
00\right\rangle $ is enhanced due to the presence of $\gamma$ (also see
Fig.4). In addition, for a certain larger regime of $\lambda$, the maximum
value of $\mathcal{B}$ becomes smaller with the increase of $\gamma$.

On the other hand, from the expression we see that the degree of nonlocality
depends on the coherent amplitude $J$, and on the squeezed parameters
$\lambda$ and $\gamma$, and on the phases $\varphi$, $\theta$. We have plotted
the Bell function $\mathcal{B}$\ as a function of the phases $\varphi$,
$\theta$ and with fixed $J=0.01$ and several different $\lambda,\gamma$, as
shown in Fig. 5. One can clearly see from Fig. 5 that $\mathcal{B}$ is always
greater than zero and for a smaller $\lambda$, the variable $\mathcal{B}$
reaches its maxmum value for $\varphi=0,\theta=\pi$ or $\varphi=\pi,\theta=0$
(see Fig.5(a),(b)); while for a larger $\lambda$, the phases $\varphi$,
$\theta$ corresponding to the maxmum value of $\mathcal{B}$ are different from
those above and vary as $\gamma$ parameter.

\begin{figure}[ptb]
\label{Fig5} \centering\includegraphics[width=12cm]{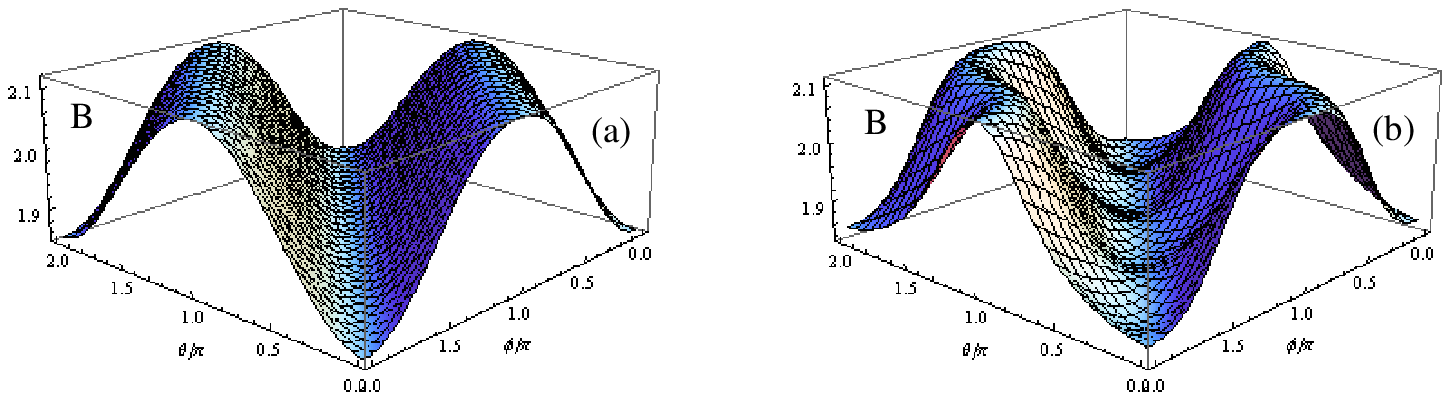}
\centering\includegraphics[width=12cm]{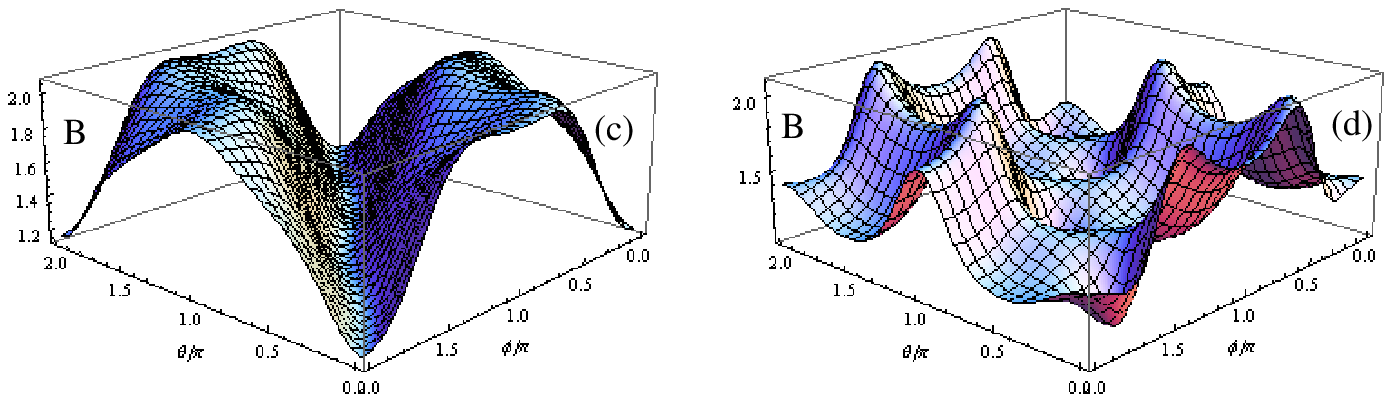}\caption{{\protect\small (Color
online) Plot of the Bell function }$\mathcal{B}$ {\protect\small as a function
of parameters }$\theta$ {\protect\small and }$\varphi$,{\protect\small for
given }$J=0.01,${\protect\small and (a) }$\lambda=0.5${\protect\small ,
}$\gamma=1;${\protect\small (b) }$\lambda=0.5${\protect\small , }$\gamma
=2;${\protect\small (c) }$\lambda=1${\protect\small , }$\gamma=1;$%
{\protect\small (a) }$\lambda=1${\protect\small , }$\gamma=2.$}%
\end{figure}

\section{Application to quantum teleportation with $V\left\vert
00\right\rangle $}

In quantum teleportation (QT), an unknown state is transmitted from a sender
(Alice) to a receiver (Bob) via a quantum channel with the aid of some
classical information. This process may be regarded as sending and extracting
quantum information via the quantum channel. QT was firstly proposed by
Barnnett et al in the discrete variable regime \cite{33} followed by
experimental demonstration \cite{34,34a}. For the continuous variables (CVs)
case, the theoretical analysis of teleportation was firstly made by Vaidman
\cite{35}. The role of teleportation in the CV quantum information is analyzed
in the review Ref.\cite{36}.

Recently, a CV teleportation protocol has been given in terms of the
characteristic functions (CFs) of the quantum states involved (input, source
and teleported (output) states) \cite{37}. By using the Weyl expansion of
density operator, it is shown that the CF $\chi_{out}\left(  \beta\right)  $
of the output state has a remarkably factorized form%
\begin{equation}
\chi_{out}\left(  \beta\right)  =\chi_{in}\left(  \beta\right)  \chi
_{E}\left(  \beta^{\ast},\beta\right)  , \label{40}%
\end{equation}
where $\chi_{in}\left(  \beta\right)  $ and $\chi_{E}\left(  \beta^{\ast
},\beta\right)  $ are the CFs of the input state and the entangled source,
respectively, $\chi_{in}\left(  \alpha_{1}\right)  =$tr$\left[  D_{1}\left(
\alpha_{1}\right)  \rho\right]  $ and $\chi_{E}\left(  \alpha_{1},\alpha
_{2}\right)  =$tr$\left[  D_{1}\left(  \alpha_{1}\right)  D_{2}\left(
\alpha_{2}\right)  \rho\right]  $, $D_{i}\left(  \alpha_{i}\right)  $ is the
displacement operator corresponding to mode $i$, and $\rho$ is the density
operator associated to the state.

In order to measure the success probability of a teleportation protocol, it is
convenient to use the fidelity of teleportation $\mathcal{F=}$tr$\left(
\rho_{in}\rho_{out}\right)  $, an overlap between the input state $\rho_{in}$
and the output (teleported) state $\rho_{out}$, which can, in the CF form, be
expressed as%
\begin{equation}
\mathcal{F=}\int\frac{d^{2}\eta}{\pi}\chi_{in}\left(  \eta\right)  \chi
_{out}\left(  -\eta\right)  . \label{41}%
\end{equation}
Substituting Eq.(\ref{40}) into (\ref{41}) yields
\begin{equation}
\mathcal{F=}\int\frac{d^{2}\eta}{\pi}\left\vert \chi_{in}\left(  \eta\right)
\right\vert ^{2}\chi_{E}\left(  -\eta^{\ast},-\eta\right)  . \label{42}%
\end{equation}
In the following we use Eq.(\ref{42}) to analyze the efficiency of
teleportation for $V\left\vert 00\right\rangle $ as a quantum channel.

Let us first consider Braunstein and Kimble protocol \cite{38} of QT for
single-mode coherent states $\left\vert \beta\right\rangle $, whose CF reads%
\begin{equation}
\chi_{\text{coh}}\left(  \alpha\right)  =\exp\left[  -\frac{1}{2}\left\vert
\alpha\right\vert ^{2}+\alpha\beta^{\ast}-\alpha^{\ast}\beta\right]  .
\label{43}%
\end{equation}
For the OTCSS, its CF is given by (see Appendix B)
\begin{equation}
\chi\left(  \alpha;\beta\right)  =\exp\left[  -\frac{1}{2}\left(
\begin{array}
[c]{cccc}%
\alpha^{\ast} & \alpha & \beta^{\ast} & \beta
\end{array}
\right)  \frac{1}{4}M\left(
\begin{array}
[c]{cccc}%
\alpha^{\ast} & \alpha & \beta^{\ast} & \beta
\end{array}
\right)  ^{T}\right]  , \label{44}%
\end{equation}
where $M$ is defined in Eq.(\ref{37c}). Upon substituting Eqs.(\ref{43}) and
(\ref{44}) into (\ref{42}), we worked out the fidelity for teleporting a
coherent state based on the OTCSS (\ref{15}),%
\begin{equation}
\mathcal{F}=\frac{1}{1-f}, \label{45}%
\end{equation}
where $f=\cosh\gamma\sinh2\lambda-\cosh^{2}\lambda-\cosh2\gamma\sinh
^{2}\lambda.$ Eq.(\ref{45}) indicates that the fidelity is only dependent on
the parameters $\lambda$ and $\gamma$, and is independent of amplitude of the
coherent state. In particular, when $\gamma=0$, Eq.(\ref{45}) just reduces to
$\mathcal{F=}(1+\tanh\lambda)/2$ \cite{39}.

Next we consider to teleport the single-mode squeezed vacuum state,
$\exp\left[  r/2\left(  a^{2}-a^{\dag2}\right)  \right]  \left\vert
0\right\rangle ,$ whose CF reads
\begin{equation}
\chi_{sq}\left(  \alpha\right)  =\exp\left[  -\frac{1}{2}\left\vert
\alpha\right\vert ^{2}\cosh2r-\frac{1}{4}\left(  \alpha^{2}+\alpha^{\ast
2}\right)  \sinh2r\right]  , \label{46}%
\end{equation}
substituting Eqs.(\ref{46}),(\ref{44}) into Eq.(\ref{42}) yields the
fidelity,
\begin{equation}
\mathcal{F}\left(  r\right)  \mathcal{=}\frac{1}{\sqrt{\allowbreak
f^{2}-2f\cosh2r+1}}. \label{47}%
\end{equation}
In order to compare the fidelity obtained by the TSVS channel and the OTCSS
channel, we plot figure for the difference fidelity ($\mathcal{F}\left(
r\right)  -\mathcal{F}\left(  0\right)  $) as a function of parameters
($\lambda$ and $\gamma$) in Fig.6. From Fig.6, one can see that there is a
region spanned by $\lambda$ and $\gamma$ in which the fidelity of OTCSS
channel is larger than the other one and the difference fidelity becomes
smaller with the increase of $r$.

\begin{figure}[ptb]
\label{Fig6}
\centering\includegraphics[width=12cm]{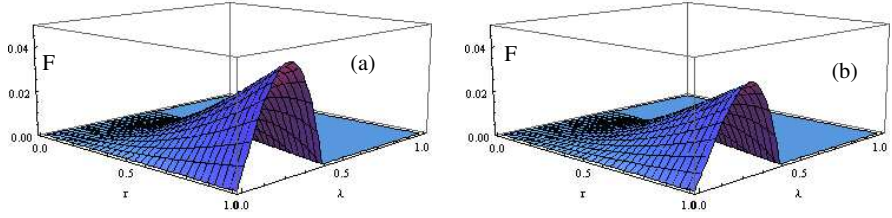}\caption{{\protect\small (Color
online) Plot of the fidelity }$\mathcal{F}$ {\protect\small as a function of
parameters }$\lambda$ {\protect\small and }$\gamma$,{\protect\small (a) the
initial coherent state, with }$r=0;${\protect\small (b) the initial squeezed
vacuum state with }$r=1.$}%
\end{figure}

\section{Conclusion}

In conclusion, we have introduced a one- and two-mode combination squeezed
state (OTCSS) which can be considered as a generalized two-mode squeezed state
and investigated the entanglement properties and quantum noncocality of this
state in terms of logarithmic negativity and the Bell's inequality,
respectively. It is shown that this state presents larger entanglement than
that of the usual two-mode squeezed vacuum state (TSVS); In a certain smaller
regime of $\lambda$, it is found that the violation of Bell's inequality
becomes more observable with increasing $\gamma$, which indicates that the
nonlocality of $V\left\vert 00\right\rangle $ can be stronger than that of
TSVS due to the presence of $\gamma.$ In addition, application to quantum
teleportation with OTCSS is also considered, which shows that there is a
region spanned by $\lambda$ and $\gamma$ in which the fidelity of OTCSS
channel is larger than that of TSVS.

\bigskip

\textbf{ACKNOWLEDGEMENT:} Work supported by a grant from the Key Programs
Foundation of Ministry of Education of China (No. ) and the Research
Foundation of the Education Department of Jiangxi Province of China (No. GJJ10097).

\textbf{Appendix A: Derivation of Eq.(\ref{25})}

Using the order-invariance of Weyl ordered operators under similar
transformations (\ref{23}) and Eqs.(\ref{7})-(\ref{8}), we have%
\begin{align}
U  &  =V^{-1}%
\genfrac{}{}{0pt}{}{:}{:}%
\delta\left(  q_{1}-Q_{1}\right)  \delta\left(  p_{1}-P_{1}\right)
\delta\left(  q_{2}-Q_{2}\right)  \delta\left(  p_{2}-P_{2}\right)
\genfrac{}{}{0pt}{}{:}{:}%
V\nonumber\\
&  =%
\genfrac{}{}{0pt}{}{:}{:}%
\delta\left(  q_{1}-Q_{1}\cosh\lambda-Q_{2}e^{-\gamma}\sinh\lambda\right)
\delta\left(  q_{2}-Q_{2}\cosh\lambda-Q_{1}e^{\gamma}\sinh\lambda\right)
\nonumber\\
&  \times\delta\left(  p_{1}-P_{1}\cosh\lambda+P_{2}e^{\gamma}\sinh
\lambda\right)  \delta\left(  p_{2}-P_{2}\cosh\lambda+P_{1}e^{-\gamma}%
\sinh\lambda\right)
\genfrac{}{}{0pt}{}{:}{:}%
\nonumber\\
&  =%
\genfrac{}{}{0pt}{}{:}{:}%
\delta\left(  \left(
\begin{array}
[c]{c}%
q_{1}\\
q_{2}%
\end{array}
\right)  -\left(
\begin{array}
[c]{cc}%
\cosh\lambda & e^{-\gamma}\sinh\lambda\\
e^{\gamma}\sinh\lambda & \cosh\lambda
\end{array}
\right)  \left(
\begin{array}
[c]{c}%
Q_{1}\\
Q_{2}%
\end{array}
\right)  \right) \nonumber\\
&  \times\delta\left(  \left(
\begin{array}
[c]{c}%
p_{1}\\
p_{2}%
\end{array}
\right)  -\left(
\begin{array}
[c]{cc}%
\cosh\lambda & -e^{\gamma}\sinh\lambda\\
-e^{-\gamma}\sinh\lambda & \cosh\lambda
\end{array}
\right)  \left(
\begin{array}
[c]{c}%
P_{1}\\
P_{2}%
\end{array}
\right)  \right)
\genfrac{}{}{0pt}{}{:}{:}%
\nonumber\\
&  =%
\genfrac{}{}{0pt}{}{:}{:}%
\delta\left(  \left(
\begin{array}
[c]{cc}%
\cosh\lambda & -e^{-\gamma}\sinh\lambda\\
-e^{\gamma}\sinh\lambda & \cosh\lambda
\end{array}
\right)  \left(
\begin{array}
[c]{c}%
q_{1}\\
q_{2}%
\end{array}
\right)  -\left(
\begin{array}
[c]{c}%
Q_{1}\\
Q_{2}%
\end{array}
\right)  \right) \nonumber\\
&  \times\delta\left(  \left(
\begin{array}
[c]{cc}%
\cosh\lambda & e^{\gamma}\sinh\lambda\\
e^{-\gamma}\sinh\lambda & \cosh\lambda
\end{array}
\right)  \left(
\begin{array}
[c]{c}%
p_{1}\\
p_{2}%
\end{array}
\right)  -\left(
\begin{array}
[c]{c}%
P_{1}\\
P_{2}%
\end{array}
\right)  \right)
\genfrac{}{}{0pt}{}{:}{:}%
, \tag{A1}%
\end{align}
which indicates that$\allowbreak$ (comparing with Eq.(\ref{20}))%
\begin{align}
U  &  =%
\genfrac{}{}{0pt}{}{:}{:}%
\delta\left(  q_{1}\cosh\lambda-q_{2}e^{-\gamma}\sinh\lambda-Q_{1}\right)
\delta\left(  p_{1}\cosh\lambda+p_{2}e^{\gamma}\sinh\lambda-P_{1}\right)
\nonumber\\
&  \times\delta\left(  q_{2}\cosh\lambda-q_{1}e^{\gamma}\sinh\lambda
-Q_{2}\right)  \delta\left(  p_{2}\cosh\lambda+p_{1}e^{-\gamma}\sinh
\lambda-P_{2}\right)
\genfrac{}{}{0pt}{}{:}{:}%
\nonumber\\
&  =\text{Eq.(\ref{25})}. \tag{A2}%
\end{align}
Thus the Wigner function of $V\left\vert 00\right\rangle $ is
\begin{align}
W\left(  q_{1},p_{1};q_{2},p_{2}\right)   &  =\frac{1}{\pi^{2}}\exp\left\{
-\left(  q_{1}\cosh\lambda-q_{2}e^{-\gamma}\sinh\lambda\right)  ^{2}-\left(
q_{2}\cosh\lambda-q_{1}e^{\gamma}\sinh\lambda\right)  ^{2}\right. \nonumber\\
&  \left.  -\left(  p_{1}\cosh\lambda+p_{2}e^{\gamma}\sinh\lambda\right)
^{2}-\left(  p_{2}\cosh\lambda+p_{1}e^{-\gamma}\sinh\lambda\right)
^{2}\right\} \nonumber\\
&  =(\ref{25}). \tag{A3}%
\end{align}

\textbf{Appendix B: Derivation of characteristic function of }$V\left\vert
00\right\rangle $

For two-mode quantum state $V\left\vert 00\right\rangle $, its characteristic
function is given by%
\begin{align}
\chi\left(  q_{1},p_{1};q_{2},p_{2}\right)   &  =\mathtt{tr}\left[
V\left\vert 00\right\rangle \left\langle 00\right\vert V^{-1}D_{1}\left(
q_{1},p_{1}\right)  D_{2}\left(  q_{2},p_{2}\right)  \right] \nonumber\\
&  =\left\langle 00\right\vert V^{-1}D_{1}\left(  q_{1},p_{1}\right)
D_{2}\left(  q_{2},p_{2}\right)  V\left\vert 00\right\rangle , \tag{B1}%
\end{align}
where $D_{i}\left(  q_{i},p_{i}\right)  $ $\left(  i=1,2\right)  $ are the
displacement operators, defined by $D_{i}\left(  q_{i},p_{i}\right)
=\exp\left[  \mathtt{i}\left(  p_{i}Q_{i}-q_{i}P_{i}\right)  \right]  .$

Noticing that the Weyl ordering of $D_{i}\left(  q_{i},p_{i}\right)  $ is
itself, $D_{i}\left(  q_{i},p_{i}\right)  =%
\genfrac{}{}{0pt}{}{:}{:}%
D_{i}\left(  q_{i},p_{i}\right)
\genfrac{}{}{0pt}{}{:}{:}%
$, and that a remarkable property of the invariance of Weyl ordered operators
under similar transformations (\ref{23}), we have
\begin{align}
&  V^{-1}D_{1}\left(  q_{1},p_{1}\right)  D_{2}\left(  q_{2},p_{2}\right)
V\nonumber\\
&  =V^{-1}%
\genfrac{}{}{0pt}{}{:}{:}%
\exp\left[  i\left(  p_{1}Q_{1}-q_{1}P_{1}\right)  \right]  \exp\left[
i\left(  p_{2}Q_{2}-q_{2}P_{2}\right)  \right]
\genfrac{}{}{0pt}{}{:}{:}%
V\nonumber\\
&  =%
\genfrac{}{}{0pt}{}{:}{:}%
\exp\left[  i\left(  p_{1}\left(  Q_{1}\cosh\lambda+Q_{2}e^{-\gamma}%
\sinh\lambda\right)  -q_{1}\left(  P_{1}\cosh\lambda-P_{2}e^{\gamma}%
\sinh\lambda\right)  \right)  \right] \nonumber\\
&  \times\exp\left[  i\left(  p_{2}\left(  Q_{2}\cosh\lambda+Q_{1}e^{\gamma
}\sinh\lambda\right)  -q_{2}\left(  P_{2}\cosh\lambda-P_{1}e^{-\gamma}%
\sinh\lambda\right)  \right)  \right]
\genfrac{}{}{0pt}{}{:}{:}%
\nonumber\\
&  =%
\genfrac{}{}{0pt}{}{:}{:}%
\exp\left[  i\left(  p_{1}^{\prime}Q_{1}-q_{1}^{\prime}P_{1}\right)  \right]
\exp\left[  \allowbreak i\left(  p_{2}^{\prime}Q_{2}-q_{2}^{\prime}%
P_{2}\right)  \right]
\genfrac{}{}{0pt}{}{:}{:}%
\nonumber\\
&  =D_{1}\left(  q_{1}^{\prime},p_{1}^{\prime}\right)  D_{2}\left(
q_{2}^{\prime},p_{2}^{\prime}\right)  , \tag{B2}%
\end{align}
where%
\begin{align}
q_{1}^{\prime}  &  =q_{1}\cosh\lambda-q_{2}e^{-\gamma}\sinh\lambda,\text{
}p_{1}^{\prime}=p_{1}\cosh\lambda+p_{2}\allowbreak e^{\gamma}\sinh
\lambda,\nonumber\\
q_{2}^{\prime}  &  =q_{2}\cosh\lambda-q_{1}e^{\gamma}\sinh\lambda,\text{
}p_{2}^{\prime}=p_{2}\cosh\lambda+p_{1}e^{-\gamma}\sinh\lambda. \tag{B3}%
\end{align}
Then substituting Eqs.(B2), (B3) into Eq.(B1) we can directly obtain the CF of
$V\left\vert 00\right\rangle ,$%
\begin{align}
\chi\left(  q_{1},p_{1};q_{2},p_{2}\right)   &  =\left\langle 00\right\vert
D_{1}\left(  q_{1}^{\prime},p_{1}^{\prime}\right)  D_{2}\left(  q_{2}^{\prime
},p_{2}^{\prime}\right)  \left\vert 00\right\rangle \nonumber\\
&  =\exp\left[  -\frac{1}{2}\left(
\begin{array}
[c]{cccc}%
q_{1} & p_{1} & q_{2} & p_{2}%
\end{array}
\right)  \sigma\left(
\begin{array}
[c]{cccc}%
q_{1} & p_{1} & q_{2} & p_{2}%
\end{array}
\right)  ^{T}\right]  , \tag{B4}%
\end{align}
or%
\begin{equation}
\chi\left(  \alpha;\beta\right)  =\exp\left[  -\frac{1}{2}\left(
\begin{array}
[c]{cccc}%
\alpha^{\ast} & \alpha & \beta^{\ast} & \beta
\end{array}
\right)  \bar{\sigma}\left(
\begin{array}
[c]{cccc}%
\alpha^{\ast} & \alpha & \beta^{\ast} & \beta
\end{array}
\right)  ^{T}\right]  , \tag{B5}%
\end{equation}
where%
\begin{equation}
\sigma=\frac{1}{2}\left(
\begin{array}
[c]{cccc}%
m_{1} & 0 & -m_{3} & 0\\
0 & m_{2} & 0 & m_{3}\\
-m_{3} & 0 & m_{2} & 0\\
0 & m_{3} & 0 & m_{1}%
\end{array}
\right)  ,\bar{\sigma}=N\sigma N^{T}=\frac{1}{4}M. \tag{B6}%
\end{equation}
When $\gamma=0,$ $\left(  m_{1}+m_{2}\right)  \rightarrow2\cosh2\lambda
,m_{3}\rightarrow\sinh2\lambda,$we have%
\begin{equation}
\chi\left(  \alpha;\beta\right)  =\exp\left[  -\frac{1}{2}\allowbreak\left(
\left\vert \alpha\right\vert ^{2}+\left\vert \beta\right\vert ^{2}\right)
\cosh2\lambda+\frac{1}{2}\left(  \alpha^{\ast}\beta^{\ast}+\alpha\beta\right)
\sinh2\lambda\allowbreak\right]  , \tag{B7}%
\end{equation}
which is just the CF of the usual TSVS.


\begin{thebibliography}{99}                                                                                               %


\bibitem {1}D. Bouwmeester et al., \emph{The Physics of Quantum Information},
(Springer, Berlin) 2000.

\bibitem {2}M. A. Nielsen and I. L. Chuang, \textit{Quantum Computation and
Quantum Information} (Cambridge University Press) 2000.

\bibitem {3}V. Buzek, J. Mod. Opt. \textbf{37} (1990) 303.

\bibitem {4}R. Loudon, P. L. Knight, J. Mod. Opt. \textbf{34} (1987) 709.

\bibitem {5}V. V. Dodonov, J. Opt. B: Quantum Semiclass. Opt. \textbf{4}
(2002) R1.

\bibitem {6}Hong-yi Fan and J. R. Klauder, Phys. Rev. A \textbf{49} (1994) 704.

\bibitem {7}Hong-yi Fan and Y. Fan, Phys. Rev. A \textbf{54} (1996) 958.

\bibitem {8}Li-yun Hu and Hong-yi Fan, Europhys. Lett. \textbf{85} (2009) 60001.

\bibitem {9}A. Einstein, B. Poldolsky and N. Rosen, Phys. Rev. \textbf{47}
(1935) 777.

\bibitem {10}Li-yun Hu and Hong-yi Fan, Phys. Rev. A \textbf{80} (2009)
022115; Opt. Commun. 282 (2009) 3734.

\bibitem {11}Hong-yi Fan and Li-yun Hu, Opt. Commun. \textbf{281} (2008) 1629;
\textbf{281} (2008) 5571.

\bibitem {12}Hong-yi Fan, J Opt B: Quantum Semiclass. Opt. \textbf{5} (2003) R147

\bibitem {13}Hong-yi Fan, Phys. Rev. A \textbf{41} (1990) 1526.

\bibitem {14}E. P. Wigner, Phys. Rev. \textbf{40} (1932) 749.

\bibitem {15}R. F. O'Connell and E. P. Wigner, Phys. Lett. A \textbf{83}
(1981) 145.

\bibitem {16}W. P. Schleich, \emph{Quantum Optics in Phase Space,} Wiley-VCH,
Berlin, 2001.

\bibitem {17}Hong-yi Fan, J. Phys. A \textbf{25} (1992) 3443; Hong-yi Fan, Y.
Fan, Int. J. Mod. Phys. A \textbf{17} (2002) 701.

\bibitem {18}Hong-yi Fan, Mod. Phys. Lett. A \textbf{15} (2000) 2297.

\bibitem {19}Hong-yi Fan, Ann. Phys. (New York) \textbf{323} (2008) 500;
\textbf{323} (2008) 1502.

\bibitem {20}G. Vidal and R. F. Werner, Phys. Rev. A \textbf{65} (2002) 032314.

\bibitem {21}J. Eisert and M. B. Plenio, J. Mod. Opt. \textbf{46} (1999) 145.

\bibitem {22}S. Virmani and M. B. Plenio, Phys. Lett. A \textbf{268} (2000) 31.

\bibitem {23}L. M. Duan, G. Giedke, J. O. Cirac, and P. Zoller, Phys. Rev.
Lett. \textbf{84} (2000) 2722.

\bibitem {24}R. Simon, Phys. Rev. Lett. \textbf{84} (2000) 2726.

\bibitem {25}J. Williamson, Am. J. Math. \textbf{58} (1936) 141; V. I. Arnold,
Mathematical Methods of Classical Mechanics (Springer-Verlag, New York, 1978).

\bibitem {26}A. Serafini, F. Illuminati, M. G. A. Paris, and S. De Siena,
Phys. Rev. A\textbf{ 69} (2004) 022318.

\bibitem {27}G. Adesso, A. Serafini, and F. Illuminati, Phys. Rev. A.
\textbf{70} (2004) 022318.

\bibitem {28}K. Banasek and K. Wodkiewicz, Phys. Rev. A. \textbf{58} (1998) 4345.

\bibitem {29}K. Banasek and K. Wodkiewicz, Phys. Rev. Lett. \textbf{76} (1996)
4344; S. Wallentowitz and W. Volgel, Phys. Rev. A. \textbf{53} (1996) 4528.

\bibitem {30}J. F. Clauser, M. A. Horne, A. Shimony, and R. A. Holt, Phys.
Rev. Lett. \textbf{23} (1969) 880.

\bibitem {31}S-W Lee, H. Jeong, and D. Jaksch, Phys. Rev. A. \textbf{80}
(2009) 022104.

\bibitem {32}J. F. Clauser, and M. A. Horne, Phys. Rev. D. \textbf{10} (1974) 526.

\bibitem {33}C. H. Bennett, G. Brassard, C. Crepeau, R. Jozsa, A. Peres, and
W. K. Wootters, Phys. Rev. Lett. \textbf{70} (1993) 1895.

\bibitem {34}D. Bouwmeester, J. W. Pan, K. Mattle, M. Eibl, H.Weinfurther, and
A. Zeilinger, Nature (London) \textbf{390} (1997) 575.

\bibitem {34a}D. Boschi, S. Branca, F. De Martini, L. Hardy, and S. Popescu,
Phys. Rev. Lett. \textbf{80} (1998) 1121.

\bibitem {35}L. Vaidmain, Phys. Rev. A \textbf{49} (1994) 1473.

\bibitem {36}S. L. Braunstein and P. van Loock, Rev. Mod. Phys. \textbf{77}
(2005) 513.

\bibitem {37}P. Marian and T. A. Marian, Phys. Rev. A. \textbf{74} (2006) 042306.

\bibitem {38}S. L. Braunstein and H. J. Kimble, Phys. Rev. Lett. \textbf{80}
(1998) 869.

\bibitem {39}Y. Yang, and F-L. Li, Phys. Rev. A. \textbf{80} (2009) 022315.
\end{thebibliography}
\end{document}